\documentclass[10pt,british,aps,prl,twocolumn]{revtex4-1}
\usepackage{color}
\usepackage{amsmath}
\usepackage{amssymb}
\usepackage{graphicx}
\usepackage[unicode=true,pdfusetitle,bookmarks=true,bookmarksnumbered=false,bookmarksopen=false,
breaklinks=false,pdfborder={0 0 1},backref=false,colorlinks=false]{hyperref}

\begin{document}
\title{Sharp reversals of steady-state nuclear polarisation as a tool for quantum sensing}

\author{Q. Chen${}^{1,2}$, Z.-Y. Wang${}^{1}$ , B. Tratzmiller${}^{1}$ , I. Schwartz${}^{1,3}$  and M.B. Plenio${}^{1}$  }
\address{${}^{1}$ Institut f\"{u}r Theoretische Physik \& IQST, Albert-Einstein Allee 11,
Universit\"{a}t Ulm, D-89081 Ulm, Germany\\
${}^{2}$Key Laboratory of Low-Dimensional Quantum Structures and Quantum Control of Ministry of Education, Department of Physics and Synergetic Innovation Center for Quantum Effects and Applications, Hunan Normal University, Changsha 410081, China \\
${}^{3}$ NVision Imaging Technologies GmbH, Albert-Einstein Allee 11,
Universit\"{a}t Ulm, D-89081 Ulm, Germany\\
}

\begin{abstract}
The polarisation dynamics of nuclear spins weakly coupled to an NV center is highly sensitive to 
the parameters of the microwave control and the nuclear Larmor frequency. What is commonly regarded as a
challenge, we propose here as a resource for quantum sensing. By varying a single experimental parameter
in a suitable set-up, i.e., the Rabi frequency of a continual microwave driving or the nuclear Larmor 
frequency, we predict periodic reversals of the steady-state polarisation of the nuclear spin. Crucially, 
interference between the transverse and longitudinal dipolar interaction of electron and nuclear spins
results in remarkably sharp steady-state polarisation reversals of nuclear spins within only a few tens 
of Hz change in the nuclear Larmor frequency. Our method is particularly robust against imperfections
such as decoherence of the electron spin and the frequency resolution of the sensor is not limited by the 
coherence time $T_2$ of the electron sensor.
\end{abstract}

\maketitle

\emph{Introduction ---}
The negatively charged nitrogen-vacancy (NV) defect center in diamond has been studied extensively over 
the past decade for nanoscale sensing \cite{wu2016diamond,Degen2017} and, more recently, as a source for 
nuclear spin hyperpolarisation. Both applications benefit from the ability to prepare the electron spin 
of the NV sensor in a pure state by short, microsecond long, laser pulses which achieve over 95\% of electron 
spin polarisation. Microwave control schemes can then transfer this electron spin polarisation to a nearby 
ensemble of nuclear spins even at ambient condition
\cite{Cai2013,london2013detecting,alvarez2015local,king2015,chen2015optical,scheuer2016optically,scheuer2017robust,FernandezAcebal2018,Ajoy2018}.

Many dynamic nuclear polarization (DNP) protocols have been developed and applied over the past several 
decades, starting from continuous microwave irradiation \cite{Abragam1958} to more efficient pulsed schemes \cite{Henstra1988a,Henstra1988b}. Common to these continuous wave protocols is the use of a long microwave pulse 
to match the Larmor frequency of the nuclear spins to the electronic Rabi rotation in the frame of reference 
of the microwave drive, a condition that is known as a Hartmann-Hahn resonance \cite{HartmannHahn}. It has been 
noticed that, owing to the weak electron-nuclear interaction, even a small detuning from this resonance can 
have a significant effect on the polarization transfer dynamics. This renders these schemes strongly dependent 
on the intensity and frequency of the microwave drive as well as the nuclear Larmor frequency. Considerable 
efforts are being expended to address this challenge for example with the development of polarisation techniques 
that are less sensitive to precise resonance conditions \cite{Schwartz2018}.

In this work we are adopting a radically different point of view and instead of regarding this strong parameter 
sensitivity of the polarisation dynamics as a challenge we will explore it as a resource for novel nuclear spin 
sensing schemes. To this end we investigate the steady-state polarisation in the off-resonant case in which the 
microwave Rabi frequency is far detuned from the nuclear Larmor frequency. In this regime we observe reversals 
of the steady-state nuclear spin polarisation that are periodic in the applied Rabi frequency of the microwave 
drive or the nuclear Larmor frequency. Crucially, we find remarkably sharp steady-state nuclear spin polarisation 
reversals and build-ups within only a few tens of Hz change in the nuclear Larmor frequency. These sharp polarisation 
reversals are induced by the interference between the transverse and longitudinal dipolar coupling components 
of electron and nuclear spins, can be applied for quantum sensing and useful for atomic-scale nuclear spin 
imaging \cite{KostCP2015,ScheuerSK+2015,ajoy2015atomic}.

In sensing applications at room temperature relaxation and decoherence processes of the NV electron spin 
typically limit spectral resolution and sensitivity in protocols that are based on long Ramsay sequences. 
In such situations a quantum memory needs to be used for improving the resolution \cite{laraoui2013high,greiner2015indirect,wang2016delayed,zaiser2016enhancing,pfender2016nonvolatile,rosskopf2016quantum}
which is experimentally challenging. The sensing scheme that we propose here does not rely on direct measurements 
of the nuclear Larmor frequency in a Ramsay setup but rather on population measurements of the nuclear spin in 
steady state. As a consequence the coherence time of the NV center does not limit the spectral resolution in our 
set-up.  

\begin{figure}
\center
\includegraphics[width=0.95\columnwidth]{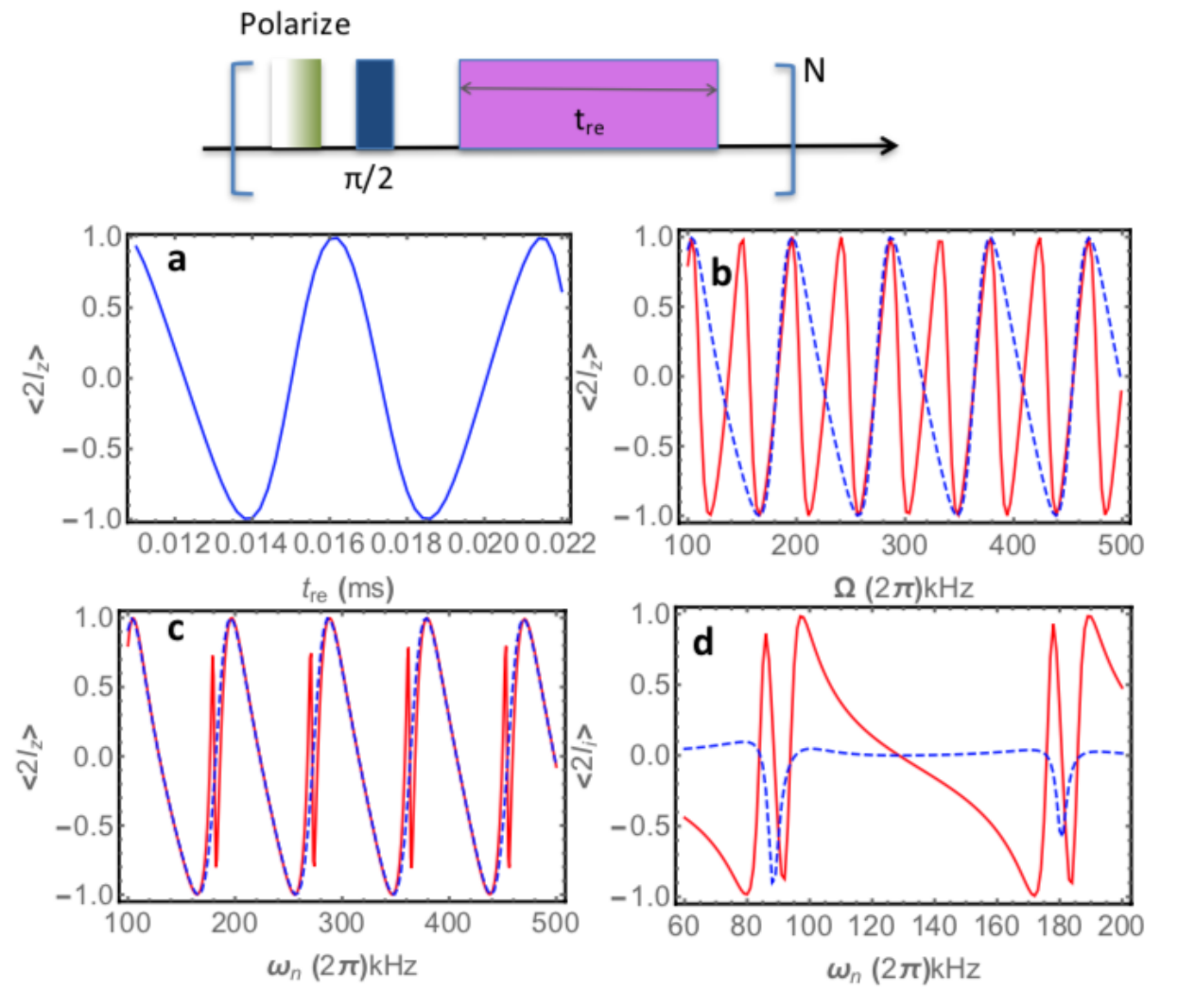}
\caption{The model consists of an NV spin interacting with a nuclear spin. The coupling parameter is $(a_{\perp},a_{\parallel})=(2\pi)(40,10)$ kHz. The nuclear spin is initially in a fully mixed state $\rho_n=\frac{\bm {I}}{2}$ and the total evolution time is 11 ms. Up: Pulse sequences in our scheme. Bottom: (a) The steady-state polarisation changes with the reset time, $\omega_n=(2\pi)$15 kHz, $\Omega=(2\pi)$200 kHz. (b) The steady-state polarisation changes with the Rabi frequency of the driving, $\omega_n=(2\pi)$15 kHz. the blue line presents $ t_{\text{re}}=11$ $\mu$s and the red dashed line shows $ t_{\text{re}}=22$ $\mu$s. (c) Periodical steady-state polarisation shows up as well as the sharp reversals with the change of nuclear Larmor frequency by the red curve, with $\Omega=(2\pi)$15 kHz. The blue dashed curve presents the case when $a_{\parallel}=0$ kHz. (d) The steady-state polarisation reversals in $z$-direction is accompanied by steady-state polarisation built-ups in $x$-direction, which is induced by the interference between the transverse and the longitudinal dipolar coupling of the electron and nuclear spins. The red solid line is as the same as in (c) and the blue dashed line shows steady-state polarisation $\langle 2I_x\rangle$. }
\label{1spin}
\end{figure}
\begin{figure*} \center
\includegraphics[width=6.3 in]{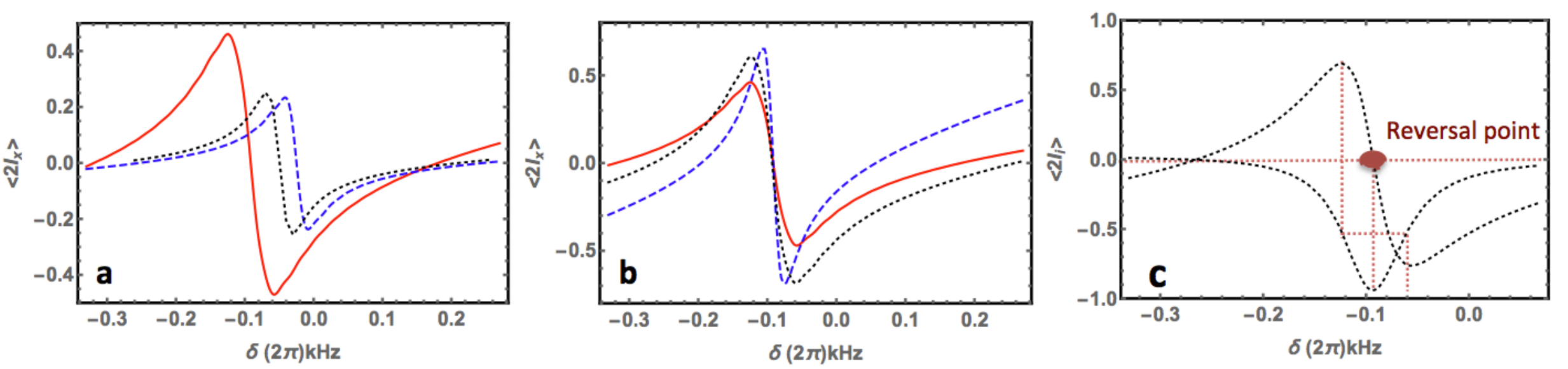}
\caption{The steady-state nuclear spin polarisation in $z$-direction $\langle 2I_z\rangle$ are shown near the polarisation reversal point. (a) The shift of the reversal pint. The red line presents the case when NV resets are applied every $ t_{\text{re}}=44$ $\mu$s and the total evolution time is 44 ms with $\Omega=(2\pi)$1 kHz and and $(a_{\perp},a_{\parallel})=(2\pi)$(4,0.5) kHz. The blue dashed line is different from the red one with $a_{\perp}=(2\pi)$2 kHz, while the black dotted line with $ t_{\text{re}}=22$ $\mu$s. (b) The change of the linewidth. The red line is as the same as in (a). The other colours are different from the red one with $\Omega=(2\pi)$2 kHz (the black dotted line), $a_{\parallel}=(2\pi)$0.25 kHz (the blue dashed line). (c) One of the black dot-dashed line is the same in (b) and the other one shows the nuclear polarisation of the steady state in $x$-direction $\langle 2I_x\rangle$. }
\label{fig2}
\end{figure*}

\emph{The model ---}
We start by considering an isolated spin pair formed by an NV center and a nuclear spin $i$ with gyromagnetic ratio  $\gamma_{n}$.
Their interaction can be described by the dipole-dipole coupling $H_{\rm{int}} = S_z \cdot \vec{A}_i\cdot\vec{I}_i$,
where $\vec{A}_i=(a_{\parallel_i}, a_{\perp_i})$ is the hyperfine vector with $a_{\parallel_i}$ and $a_{\perp_i}$ denotes the related coupling components parallel and perpendicular to the nuclear spin quantisation axes. In an external magnetic field $B_0$, the effective Lamor frequency of nuclear spin is
$\omega_n = \gamma_{n}B_0-\frac{a_{\parallel i}}{2}$. A microwave (MW) field (Rabi frequency
$\Omega_{mw}$ and frequency $\omega$) is applied to one specific electronic transition of the NV center, i.e.., $|m_s=0\rangle \leftrightarrow |m_s=-1\rangle$ transition. Within the subspace $\{|m_s=0\rangle, |m_s=-1\rangle\}$, the effective Hamiltonian of the NV spin is $H_{NV}=\Omega\sigma_z$, and
the microwave dressed states $\{|+_x\rangle=\frac{1}{\sqrt{2}}(|0\rangle+|-1\rangle), |-_x\rangle=\frac{1}{\sqrt{2}}
(|0\rangle-|-1\rangle)\}$ define $\sigma_z=\frac{1}{2}(|+_x\rangle\langle+_x|-|-_x\rangle\langle-_x|)$. We assume $\hbar=1$ as customary and the Hamiltonian of the model is given by
\begin{eqnarray}%
    H'_{tot}&=&\Omega\sigma_z
    +\omega_nI^{z}_i+\sigma_x(a_{\parallel_ i} I^{z}_i+a_{\perp_i}I^{\hat{x}}_{i}),
\label{Htot}
\end{eqnarray}
one can go to rotating frame with $H_0=\Omega\sigma_z+\omega_nI^{z}_i$ and
\begin{eqnarray}%
    H_{in}&=& a_{\perp_i}(e^{i \Delta_+t}\sigma_+ I^{+}_i+e^{i \Delta_-t} \sigma_+ I^{-}_i) \\ \nonumber & &+a_{\parallel_ i} e^{i\Omega t}I^{z}_i \sigma_++H.C.,
\label{Htot''}
\end{eqnarray}
in which
$\Delta_\pm=\Omega\pm\omega_n$.
%

\begin{figure*}
\center
\includegraphics[width=6.3 in]{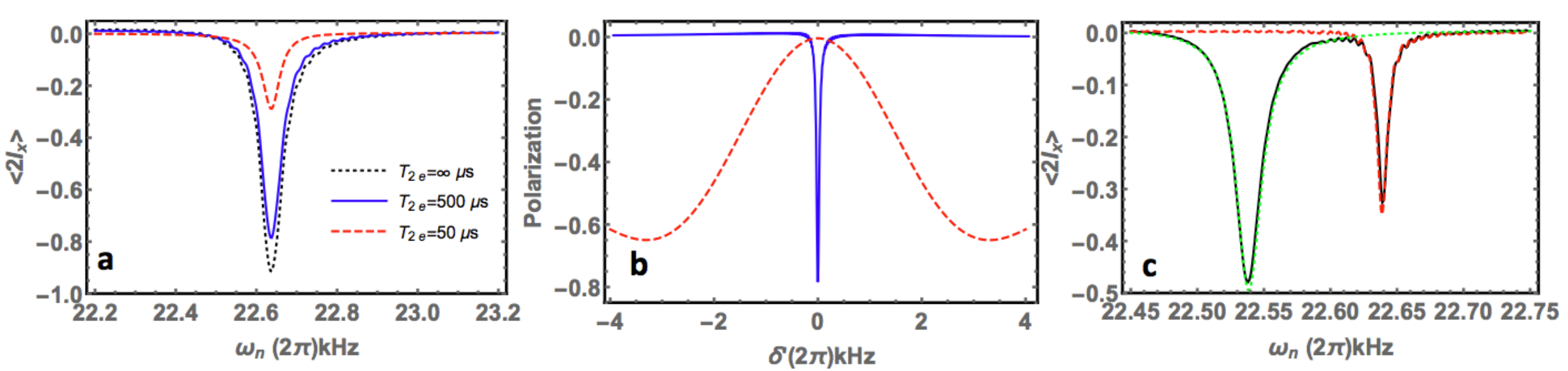}
\caption{(a) The effect of decoherence time of NV centers. NV resets are applied every $ t_{\text{re}}=44$ $\mu$s and $T_{ev}=44$ ms with $\Omega=(2\pi)$2 kHz, when the NV interacts with a nuclear spin $(a_{\perp},a_{\parallel})=(2\pi)$(4,0.5) kHz. (b) The comparison between our scheme and the direct polarisation leakage scheme. The blue line is as the same as in (a). The red dashed line presents the direct sensing through polarisation leakage of the NV center $\langle 2\sigma_z\rangle$ (initial polarisation of the NV center $\langle 2\sigma_z\rangle=-1$) with the decoherence time of the NV center is given by $T_{2e}=500$ $\mu$s and evolution time 250 $\mu$s, the Rabi frequency of the MW driving is equal to the nuclear Larmor frequency. $\delta'$ is the shift from their resonant nuclear Lamor frequencies. (c) Applications of the method to high-resolution spectroscopy with the sensor coupled to two closed nuclear spin with slightly different frequencies. Both frequency components can be clearly achieved in the resulting spectrum even though the frequencies are only 50 Hz apart, $T_{2e}=500$ $\mu$s and $\omega_n = \gamma_{n}B_0-\frac{a_{\parallel 1}}{2}$. The black solid line represents the total signal of the two spins, when the coloured lines show individual contributions. The couplings are given as $(a_{\perp_1},a_{\parallel_1},a_{\perp_2},a_{\parallel_2})=(2\pi)$(4,0.1,5,0.2) kHz, the blue line presents $\langle 2I_x\rangle$. The evolution time is 320 ms. }
\label{3spin}
\end{figure*}

The NV spin is initialised to the ground state by green laser illumination and transferred to state $|-_x\rangle$ by using a microwave-$\pi/2$ pulse. We follow the basic cycles for nuclear spin polarisation, namely an iteration between evolution according to Hamiltonian (\ref{Htot}) followed by reinitialisation of the electron spin to $|-_x\rangle$. The density matrix of the system evolves according to
\begin{eqnarray}%
    \rho_n(t+ t_{\text{re}}) \rightarrow  \mathrm{Tr_e}  [U( t_{\text{re}})(\rho_n (t)\otimes  |-_x\rangle \langle -_x|)U^{\dag}( t_{\text{re}})]
     \label{evol}
\end{eqnarray}%
$ \mathrm{Tr_e}$ presents the trace over the electron and $\rho_n$ is the density matrix of nuclear spins in the system.
All the numerical simulations are implemented by using Eq. (\ref{evol}), the reset of the NV to the state $|-_x\rangle$ every $ t_{\text{re}}$ introduces an effective interaction time in each cycle and $U( t_{\text{re}})=e^{-iH'_{tot} t_{\text{re}}} $.

When $\omega_n\sim2k\pi/ t_{\text{re}}$ ($k=1,2,...$), one can do an expansion on a detuning $\delta=\omega_n-2k\pi/ t_{\text{re}}$ ($\delta t_{\text{re}}\ll 1$ and $k=1$ is chosen for simplification), the time evolution operator is given by $U_{ t_{\text{re}}}=U_0U_{\rm{int}}=e^{-iH_0 t_{\text{re}}} \mathbb{T}e^{-\int_0^{ t_{\text{re}}}[iH_{in} dt]}$ with $e^{-iH_0 t_{\text{re}}}=e^{-i t_{\text{re}}(\Omega\sigma_z+\delta I^{z}_i)}$.  According to the second-order expansion, the master equation is given by
\begin{eqnarray}%
\rho_n(t+ t_{\text{re}})=\sum_{j,k=\pm}[C_z I^{z}_i+C_x I^{x}_i+C_y I^{y}_i,\rho_n(t)]+\nonumber
\\
 (\mathcal{D}[I_{j}]+\mathcal{M}_{j}[I_k]+\mathcal{M}[I_z])\rho_n(t),
    \end{eqnarray}
in which $\mathcal{D}[I_{\pm}]\rho_n=\pm g_{\pm}g_{\pm}^*(I_{\mp}I_{\pm}\rho_n+\rho_nI_{\mp}I_{\pm}-2I_{\pm}\rho_nI_{\mp})$, $\mathcal{M}_{\pm}[I_+]\rho_n=g_{z}^*g_{\pm}(I_{z}I_{\pm}\rho_n+\rho_nI_{z}I_{\pm}-I_{\pm}\rho_nI_{z})$, $\mathcal{M}_{\pm}[I_-]\rho_n=2g_{z}g_{\pm}^*(I_{\mp}I_{z}\rho_n+\rho_nI_{\pm}I_{z}-2I_{z}\rho_nI_{\mp})$, $\mathcal{M}[I_z]\rho_n=-2g_zg_z^*I_z\rho_nI_z$. The detailed expansion and defined parameters in Eq. (4) are given in SM \cite{SI}.

When $|\delta|\gg |C_x|, |C_y|,....,|g_zg_z^*|$, the steady state polarisation of the system is given by $\rho^z_{ss} \simeq N'_c(\frac{1-\cos\Delta_+ t_{\text{re}}}{1-\cos\Delta_- t_{\text{re}}}|\uparrow_i\rangle\langle\uparrow_i|+|\downarrow_i\rangle\langle\downarrow_i|)$ with the nomalized coefficient $N'_c$, see SM \cite{SI}.The steady state polarisation of the nuclear spin is determined by the competition between the dissipation items $\mathcal{D}[I_{+}]$ and  $\mathcal{D}[I_{-}]$.  Therefore, by controlling the Rabi frequency of MW driving $\Omega$ (Fig. \ref{1spin}a), the reset time $ t_{\text{re}}$ or the nuclear Larmor frequency (Fig.1(a, b and c)), the effective dissipation rates are shifted and periodical reversals of the nuclear steady state polarisation are shown. Namely, the nuclear polarisation reversals arise from the imbalance between effective flip-flop rate from $\sigma_+I_-+\sigma_-I_+$ and flip-flip rate from $\sigma_+I_++\sigma_-I_-$ , which is similar to the polarisation reversals in Ref.\cite{Wang}. These also fit well with Fermi golden rules by calculating the transition probabilities between $|-,\uparrow\rangle\leftrightarrow|+,\downarrow\rangle$ (flip-flop) and $|-,\downarrow\rangle\leftrightarrow|+,\uparrow\rangle$ (flip-flip) \cite{SI} and our numerical simulations in Fig. 1(a, b and c).

\emph{Sharp polarisation reversals for quantum sensing ---}
When $\omega_n\sim2\pi/ t_{\text{re}}$ and $|\delta|$ is comparable to $|C_x|, |C_y|,....,|g_zg_z^*|$, reversals of the positive and negative steady state polarisation $\langle 2I_z\rangle$ of  $z$-direction shows up as well as the built-up of negative steady-state polarisation $x$-direction $\langle 2I_x\rangle$ around the same nuclear Larmor frequency, (see Fig. 1c and d). Additionally, there are similar characters of steady-state polarisation of $z$-direction and $x$-direction, as shown in Fig. 1d and Fig. 2c, the reversal point of steady-state polarisation $\langle 2I_z\rangle$ is coincide with the trough of the negative steady-state polarisation $\langle 2I_x\rangle$ built-up and linewidths are the same. These phenomenons arise from interference between the transverse and longitudinal dipolar interaction of electron and nuclear spins. Notice that the transversal interactions $\sigma_xa_{\parallel_ i} I^{z}_i$ does not commute with longitudinal interaction $\sigma_xa_{\perp_i}I^{\hat{x}}_{i}$, there are several interference items of $I_x$, $I_y$ and $I_z$ in the second-order expansion of the time-dependent evolution operator $\mathbb{T}e^{-\int_0^{ t_{\text{re}}}[iH''_{tot} dt]}$, as well as $\mathcal{M}_{j}[I_k]$ and $\mathcal{M}[I_z]$ in the second-order expansion of the master equation. These interference items are negligible when $|\delta|\gg |C_x|, |C_y|,....,|g_zg_z^*|$, but play key roles when $|\delta|$ is comparable to $|C_x|, |C_y|,....,|g_zg_z^*|$, details are given in SM \cite{SI}.

Through careful calculations, near the anomalous points $\omega_n\sim2\pi/ t_{\text{re}}$, the related items of the steady polarisation in $z$-direction $\langle 2I_z\rangle$ could be simplified as $ \rho_{ss}^z \simeq N_c[\frac{\delta-\frac{a_{\perp_ i}^2}{8\omega_n}+\frac{a_{\parallel}^2 t_{\text{re}}}{2}}{\delta-\frac{a_{\perp_ i}^2}{8\omega_n}-\frac{a_{\parallel}^2 t_{\text{re}}}{2}}|\downarrow_i\rangle\langle\downarrow_i|+|\uparrow_i\rangle\langle\uparrow_i|]$, in which a small detuning is defined as $\delta=\omega_n-2\pi/ t_{\text{re}}$. Therefore, the reversal points is shifted from $2\pi/ t_{\text{re}}$ by $\frac{a_{\perp_ i}^2}{8\omega_n}$ (see Fig. 2) and the linewidth of the spectra, namely the frequency resolution of the anomalous steady state polarisation reversal is approximated  to be $a_{\parallel}^2 t_{\text{re}}$, which fits the numerical simulations in Fig. 2. We take steady-state polarisation $\langle 2I_z\rangle$ in $z$-direction as an example to illustrate the characters of these reversal points, because characters of steady-state polarisation of $z$-direction and $x$-direction are quite similar.



There are several characters of these reversal points. i) They are periodically shown up via the shift of the Larmor frequency of nuclear spin and very sensitive to the shift. ii) With the given reset time of the electron spin, the shift of the reversal points from $2k\pi/ t_{\text{re}}$ depends on the transverse coupling, and the frequency linewidth of the spectra is determined by the longitudinal dipolar coupling, as shown in Fig. 2. iii) It is quite robust. The electron spin is reinitialized every tens of $\mu$s, which makes the steady polarisation buit-up not sensitive to electron spin decoherence, as shown in Fig. 3a. The scheme works when the reset time is within the decoherence time of the NV center and the linewidth is not related to the NV center decoherence.

All these features make our scheme has potential applications for quantum sensing. Specifically, we detect the steady-state polarisation signal of nuclei located in close proximity to the NV center.  The electron spin of the NV center can be optically initialized and read out by using laser illumination. One can have the nuclear spins to be polarised in the off-resonant case with the NV resets, then read out the steady-state polarisation through the NV center \cite{scheuer2017robust}. As we discussed, the obtained 1D spectra (see Fig. 3a and b) shows the information on the dipolar coupling to the nuclear spins. The shift of the reversal point from $2\pi/ t_{\text{re}}$ depends on the transverse coupling and the longitudinal dipolar coupling can be derived from the sharpness, which gives parameters $(a_{\parallel}, a_{\perp})$. 


%
\begin{figure}
\center
\includegraphics[width=3.3 in]{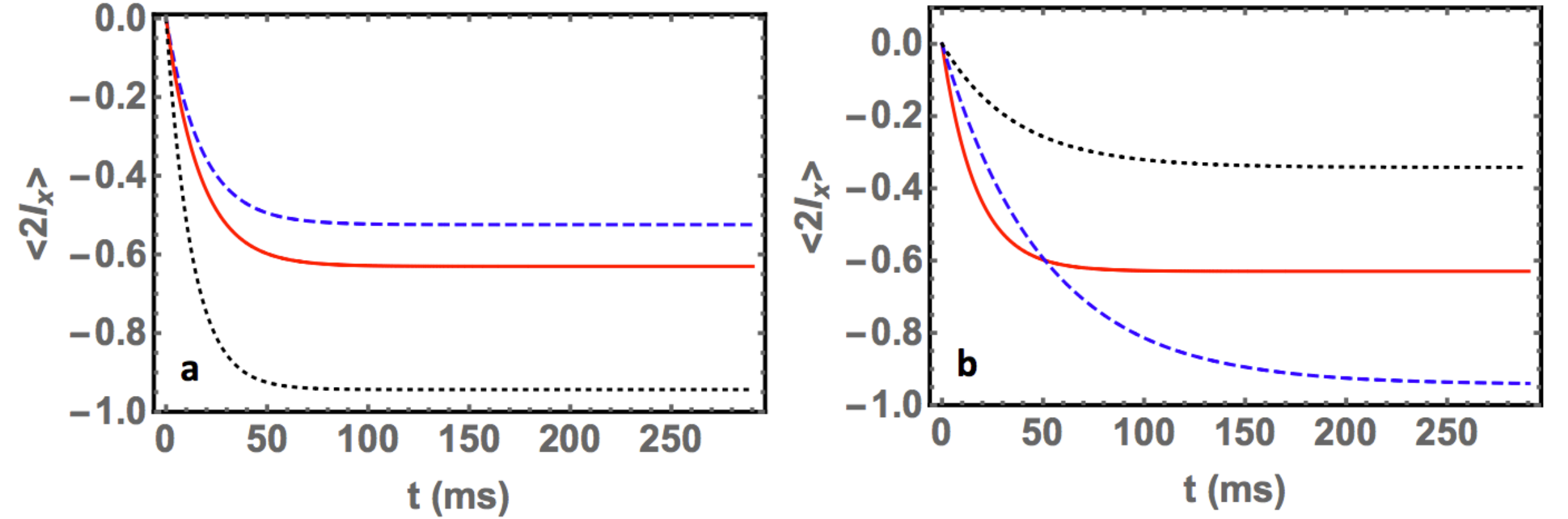}
\caption{The converging of the steady-state nuclear spin polarisation in $x$-direction $\langle 2I_x\rangle$ at the polarisation trough. The nuclear spin converges to steady-state polarisation when NV resets are applied every $t_{\text{re}}=44$ $\mu$s with $(a_{\perp},a_{\parallel})=(2\pi)$(4,0.5) kHz, $\omega_n=(2\pi)$22.64 kHz and $\Omega=(2\pi)$2 kHz (the red line). (a) The other lines are different from the red one with $\Omega=(2\pi)$ 1 kHz (the black dotted line), $a_{\perp}=(2\pi)$2 kHz (the blue dashed line). (b) The red line is the same as the one in (a). The blue dashed one is different from the red one with $a_{\parallel}=(2\pi)$0.25 kHz, while the black dotted line is with $t_{\text{re}}=22$ $\mu$s.}
\label{Fig4}
\end{figure}

\emph{Converging time, sensitivity and resolution ---}
To investigate the converging time of the steady-state polarisation, we take steady-state nuclear spin polarisation in $x$-direction $\langle 2I_x\rangle$ at the trough as an example. The converging of the system at the trough is shown in Fig. 4. The converging time of nuclear polarisation built-up is not related to the MW driving and transversal coupling strength $a_{\perp}$ (Fig. 4a), but inversely proportional to $a_{\parallel}^2 t_{\text{re}}/16$ (Fig. 4b). It is determined by $\mathcal{M}[I_z]$ and fits the calculation of the worst-case asymptotic convergence speed \cite{Viola,Riccardo}. The converging time is the largest possible chosen interrogation time of a single run. Therefore the sensitivity per averaging time of our scheme is proportional to $1/\sqrt{T_{2e}}$, see SM \cite{SI} for the details. In the traditional method of sensing through the polarisation leakage of the electron spin  \cite{Cai2}, it is limited by the decoherence time of the NV center $T_{2e}$ and proportional to $1/\sqrt{T_{2e}}$, which is as the same as the traditional method of the direct polarisation leakage scheme.

The resolution in our scheme is controllable (given by $a_{\parallel}^2 t_{\text{re}}$) and could be much more smaller than the traditional method by using the polarisation leakage of the electron spin, see Fig. 3a. However, these anomalous steady state polarisation reversal points are induced by the interference between interactions of $a_{\parallel_ i} \sigma_xI^{z}_i$ and $a_{\perp i}\sigma_xI^{\hat{x}}_{i}$, the linewidth will be limited by nuclear decoherence time $T_2$ of the target (see SM \cite{SI} for details). To demonstrate the ability of the steady-state polarisation reversal points to spectrally resolve nearby signals, we expose the sensor to two closed nuclear spins with slightly different frequencies.  As shown in Figure 3c and d, both frequency components can be clearly achieved in the resulting spectrum even though the Lamor frequencies are only 50 Hz apart. The sensing is related to the steady state of the target nuclear spin which makes the sensing is robust to electron spin decoherence.


\emph{Conclusion---} In conclusion, we show the remarkable results of periodic steady state polarisation reversals and built-ups when a nuclear spin interacts with a periodically reinitialised electron spin. Based on these steady state polarisation reversal points, we provide a new method for quantum sensing, particularly robust against imperfections such as decoherence from the electron spin, with the resolution of the sensor not being limited by the decoherence time $T_2$ of the electron sensor.

\emph{Acknowledgements ---} This work was supported by the ERC Synergy grant BioQ, the EU project HYPERDIAMOND, the
QuantERA project NanoSpin, the BMBF project DiaPol and the DFG-CRC 1279.


\end{document}